\documentclass[unnumsec,webpdf,modern,large]{mam-authoring-template}%

\graphicspath{{Fig/}}

\theoremstyle{thmstyleone}%
\theoremstyle{thmstyletwo}%
\theoremstyle{thmstylethree}%

\usepackage{hyperref}
\definecolor{linkColor}{rgb}{0.7,0,0}
\hypersetup{pdfborder={0 0 0},colorlinks=true,urlcolor=linkColor,citecolor=linkColor}

\begin{document}

\journaltitle{arXiv}
\DOI{DOI HERE}
\copyrightyear{2025}
\pubyear{2014}
\access{Advance Access Publication Date: Day Month Year}
\appnotes{Original Article}

\firstpage{1}


\title[MAPED]{Multi-angle precession electron diffraction (MAPED): a versatile approach to 4D-STEM precession}

\author[1,$\dagger$, $\ast$]{Stephanie M. Ribet}
\author[1, $\dagger$]{Rohan Dhall}
\author[2]{Colin Ophus}
\author[1, $\ast$]{Karen C. Bustillo}

\authormark{Ribet et al.}

\address[1]{\orgdiv{National Center for Electron Microscopy, Molecular Foundry}, \orgname{Lawrence Berkeley National Laboratory}, \orgaddress{\street{Berkeley}, \postcode{94720}, \state{CA}, \country{United States}}}

\address[2]{\orgdiv{Department of Materials Science and Engineering}, \orgname{Stanford University}, \orgaddress{\street{Palo Alto}, \postcode{94305}, \state{CA}, \country{United States}}}

\corresp[$\ast$]{Corresponding author. \href{email:email-id.com}{sribet@lbl.gov, kbustillo@lbl.gov}}

\received{Date}{0}{Year}
\revised{Date}{0}{Year}
\accepted{Date}{0}{Year}

\abstract{Precession of a converged beam during acquisition of a 4D-STEM dataset improves strain, orientation, and phase mapping accuracy by averaging over continuous angles of illumination.
Precession experiments usually rely on integrated systems, where automatic alignments lead to fast, high-quality results. 
The dependence of these experiments on specific hardware and software is evident even when switching to non-integrated detectors on a precession tool, as experimental set-up becomes challenging and time-consuming.
Here, we introduce multi-angle precession electron diffraction (MAPED): a method to perform electron diffraction by collecting sequential 4D-STEM scans at different incident beam tilts.
The multiple diffraction datasets are averaged together post-acquisition, resulting in a single dataset that minimizes the impact of the curvature and orientation of the Ewald sphere relative to the crystal under study. 
Our results demonstrate that even four additional tilts improved measurement of material properties, namely strain and orientation, as compared to single-tilt 4D-STEM experiments.
We show the versatility and flexibility of our MAPED approach with data collected on a number of microscopes with different hardware configurations and a variety of detectors.
}


\maketitle

\section{Introduction}

\begin{figure*}[ht]
\centering
\includegraphics[width = 0.95\textwidth]{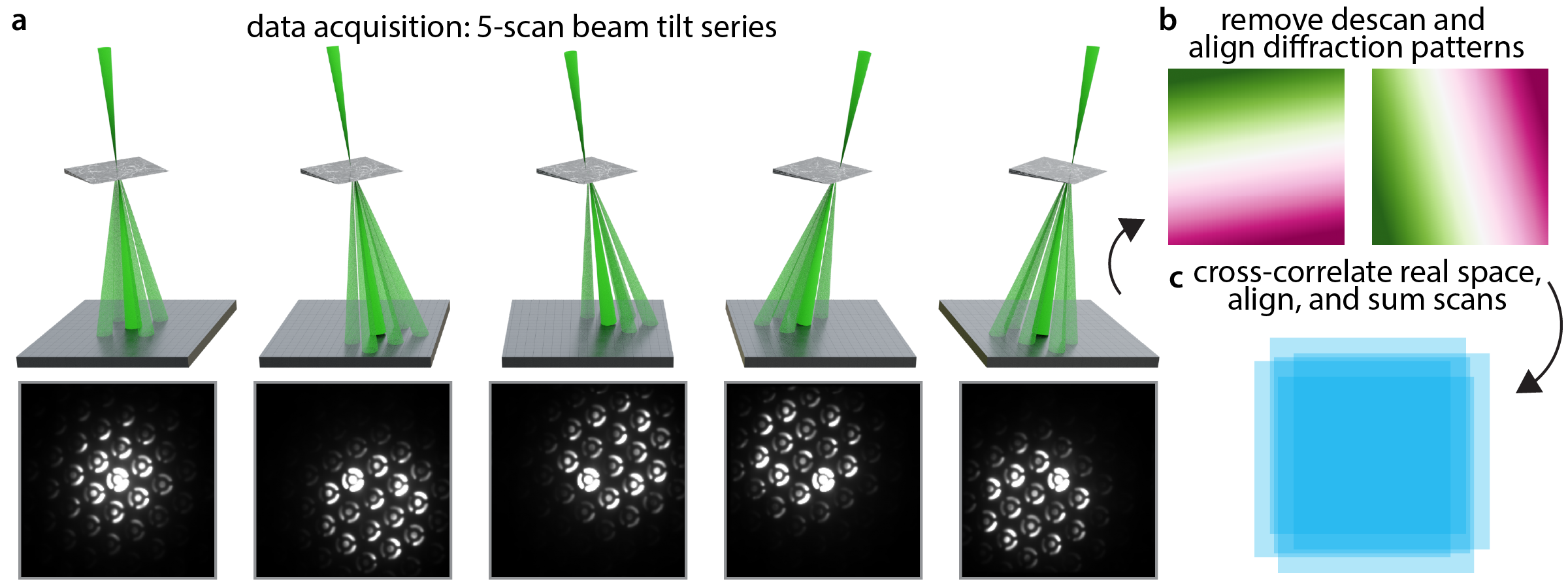}
  \caption{{\bf MAPED experimental method and processing}. (a) Diffraction patterns measured from five unique beam tilts. (b) Descan of the diffraction patterns as a function of probe position is corrected first. (c) After, real space field-of-view shifts are corrected across the five 4D-STEM experiments.}
  \label{fig1}
 \end{figure*}

The design and development of materials relies on accurate structural characterization down to the atomic scale. 
Scanning / transmission electron microscopy (S/TEM) plays an essential role in high-resolution materials characterization due to its small probe size and multimodal nature~\citep{ophus2023quantitative}.  
For crystalline materials, this characterization often includes determining the local phase and orientation~\citep{brunetti2011confirmation} and deformation strain~\citep{hytch2014observing}.
Traditionally in S/TEM experiments, strain has been mapped with real-space analysis methods ~\citep{hytch1998quantitative}. 
However, conventional imaging relies on atomic-resolution images, which limits the possible field of view, requires the crystal to be aligned in specific orientations relative to the electron beam, and needs sufficiently high electron fluence to achieve atomic resolution. 

Modern STEM experiments often include 4D-STEM, where an electron beam is rastered across a specimen, and diffraction data are collected at each position in real space creating a 4-dimensional dataset~\citep{ophus2019four}. 
The 4D-STEM data can be analyzed to map properties of materials that would be inaccessible or challenging to probe with conventional methods. 
For 4D-STEM strain experiments, data are collected with a probe formed by a small convergence angle, which reduces overlap of diffracted beams in reciprocal space. 
This approach can be extended to amorphous materials, which is not possible with conventional high resolution techniques~\citep{gammer2018local,kennedy2025mapping}.
Similarly, the orientation and phase of a sample can be studied using 4D-STEM, including automated crystal orientation mapping (ACOM) when coupled with computational diffraction pattern matching ~\citep{rauch2010automated,  cautaerts2022free, ophus2022automated}.
Despite the immense benefits of strain mapping and ACOM techniques for studying the structure and properties of material systems, there can be practical challenges for thicker and more complex structures, due to multiple scattering of the electron beam leading to dynamical effects in the diffraction patterns~\citep{muller2012strain}. 

One of the most influential innovations for mitigating dynamical diffraction effects in S/TEM was introduced by Vincent and Midgley in 1994, who demonstrated precession electron diffraction with a convergent beam~\citep{vincent1994double}. 
This approach, which builds on earlier “hollow cone” illumination strategies ~\citep{krakow1976method, kondo1984new}, uses rocking of the Ewald sphere to sample a wider range of diffraction space.
The resulting diffraction patterns exhibit more uniform Bragg disk intensities, are less sensitive to sample mistilt, and can more easily excite higher-order reflections, making them more straightforward to index and more robust against dynamical scattering~\citep{vincent1994double, rauch2010automated, rouviere2013improved}.
By averaging over multiple directions of incident electron beams, diffraction contrast can be suppressed, and it has even enabled the measurement of polarization through differential phase contrast (DPC) measurements~\citep{kohno2022development}.
Similar to this work, recent studies have implemented multiple tilts for analysis of electric polarization as an alternative to DPC imaging~\citep{toyama2024direct, flathmann2025sequential}.

Precession electron diffraction can now be directly integrated into 4D-STEM workflows, where synchronized beam precession, raster scanning, and camera acquisition are handled by vendor-supplied systems. 
These integrated tools maintain precise alignment during acquisition, producing sharp diffraction patterns, which can be leveraged for high-quality analysis  including strain mapping, orientation indexing, and digital dark field analysis~\citep{correa2024high, diebold2025template, maclaren2025digital}.
These precession tools can be challenging to align, especially when they are used with hardware such as detectors and apertures that are not fully integrated in the system. 
Although there is an increasing emphasis and availability on precession tools, few microscopes still have integrated precession equipment, making the implementation of this technique limited.

Here we present multi-angle precession electron diffraction (MAPED): a data-acquisition and post-processing strategy based on sequentially acquired 4D-STEM datasets at five distinct beam tilts to perform precession-like experiments.
This sequential tilting approach allows us to replicate the averaging effect of hardware-integrated precession with the benefit of flexibility across different microscopes and detectors. 
We found that even four additional tilts ($\sim$0.5-1$^\circ$) dramatically improved the accuracy of the resulting strain maps.
For challenging cases, such as phase discrimination in structurally similar crystals, we also simulated silicon samples along the [110] zone at various thicknesses to determine how well disk intensities matched those predicted by structure factor calculations. 
To test robustness, we examined data acquired on both corrected and uncorrected TEMs, using multiple camera systems, and assessed the effect of tilt count on indexing fidelity and strain reconstruction accuracy. 
We also provided open-source code such that this can be implemented at other microscopy centers.
This work provides a generalizable and practical framework for enhancing 4D-STEM strain and orientation measurements, with potential for broad adoption across electron microscopy facilities.

\section{Materials and Methods}

\subsection{Data acquisition}

An overview of our measurement scheme is shown in Fig.~\ref{fig1}. 
First, we collected five 4D-STEM datasets, each at a different beam tilt. 
We measured and removed descan through computational post processing. 
This ensured that when summing up diffraction data, each pixel on every diffraction pattern corresponds to the same scattering angle of the electron beam. Tilting of the electron beam can also create shifts in the image plane as shown in Fig.~\ref{fig1}c.
These shifts are handled by registration of virtual images generated from each 4D-STEM dataset.    
Once alignment of the datasets converged, we averaged them and  analyzed the average data in the same manner as conventional 4D-STEM data.

All TEM data were acquired at 300~kV, and the sample was not moved during acquisition.
4D-STEM data in Fig.~\ref{fig2} and ~\ref{fig3} were collected on the TEAM I microscope at the Molecular Foundry, a modified FEI Titan double-aberration-corrected microscope operated in nanoprobe mode.
The data were acquired on the Dectris Arina camera operating in full frame mode (192x192)~\citep{stroppa2023stem}.

The data in Fig.~\ref{fig2} were collected with a probe defined by a 10 $\mu$m custom bullseye aperture with a 2.2 mrad convergence semiangle ~\citep{zeltmann2020patterned}. 
Five scans were acquired at indicated beam tilts of $(q_x, q_y) = (0, 0)$, $(+9, 0)$, $(-9, 0)$, $(0, +9)$, and $(0, -9)$~mrad using the probe corrector (CESCOR, CEOS) beam tilt controls. 
After tilting the illumination beam, the diffraction pattern was manually shifted to the approximate center of the detector using post specimen deflectors.
We scanned 256x256 probe positions with a dwell time of 500 µs. 
The control data were collected with 2500 µs dwell time with no added beam tilt. 
The electron fluence of both the tilt-averaged 4D-STEM dataset, and the control dataset was approximately 2.1 $\times$ 10$^3$e$^{-1}$/\AA$^2$. 
Our sample was a  MAG*I*CAL calibration sample available from Ted Pella, which consists of alternating silicon and silicon-germanium layers grown on a single crystal silicon substrate, where the silicon germanium layers have composition Si$_{0.81}$Ge$_{0.19}$, according to the specifications provided from the manufacturer.
The multilayers of Si and SiGe are grown with the [100] axis perpendicular to the interface; we chose this axis to be the x-axis for the strain maps. 
EELS data were collected using the Gatan K3 camera to calculate a thickness of t$\slash \lambda$ = 1.04 (Fig.~\ref{fig_si:eels}), which corresponds to a thickness of approximately 180 nm~\citep{lee2002measurement}.

The data in Fig.~\ref{fig3} were acquired following a similar method as described for Fig.~\ref{fig2}. 
Here a 5 $\mu$m custom  aperture was used forming a probe with a 0.64 mrad convergence semiangle.
Five scans were acquired with a dwell time of 100~µs at indicated beam tilts of $(q_x, q_y) = (0, 0)$, $(+17, 0)$, $(-17, 0)$, $(0, +17)$, and $(0, -17)$~mrad using the probe corrector software.
The control data were collected with 500 µs dwell time with no added beam tilt. 
The cumulative dose of both the control dataset and the tilt averaged dataset was approximately 2.8$\times$ 10$^2$e$^{-1}$/\AA$^2$.
The data were collected on the commercially available $\sim$30 nm thick evaporated aluminum diffraction standard sample from Ted Pella, suspended on a lacey carbon grid. 

The tilts from the probe corrector software were calibrated using a large convergence angle as shown in Fig.~\ref{fig_si:team_i_cali}. 
A 30 mrad probe was defined by a 70 µm aperture. 
Using this convergence angle as a reference, the beam was tilted in $q_x$ and $q_y$.
The measured tilts were approximately linear with the beam-tilt setting of the probe corrector and the calibrated tilt was about 10\% less than the indicated tilt. 
Therefore the datasets in Fig.~\ref{fig2} were acquired with approximately a 8.1 mrad (0.46$^{\circ}$) deflection, while the datasets in Fig.~\ref{fig3} were acquired with a 15.3 mrad (0.87$^{\circ}$) deflection.
This calibration does not include a possible beam shift.

The data in Fig.~\ref{fig5} were acquired on an uncorrected Thermo Fisher Scientific ThemIS at the Molecular Foundry operated in microprobe mode, where the illumination beam was tilted  using the ``rotation center" alignment feature of the microscope. 
By drawing a calibrated circle in the diffraction “View” mode and adjusting the rotation center until the unscattered beam intersected the circle, the corresponding rotation center X and Y values were recorded. 
These values varied approximately linearly with tilt between 9 and 17.5 mrad as shown in Fig.~\ref{fig_si:Rot_cal}. 
This calibration method does not account for a beam shift, but it is a quick and approximate approach.
Data were collected using Gatan 4D-STEM software on a K2-IS camera operating at 400 frames per second. 
No diffraction shift correction was needed as the sensor was large enough to accommodate the pattern shift, making the entire workflow fast and straightforward.
4D-STEM data was also acquired on a Thermo Fisher Ceta camera using legacy TEM Imaging and Analysis (TIA) software. 
While effective, this workflow was more time-consuming, as the region of interest box had to be manually positioned and sized for each scan in the absence of scripting support. 
Diffraction patterns were acquired at 33 frames per second on the Ceta camera.

\subsection{Data processing and analysis}

As shown in Fig.~\ref{fig1}b-c, the five-tilt series for both the strain and orientation mapping datasets were aligned in reciprocal space and then real space using the following procedure. 
First, we aligned the datasets in reciprocal space in order to ensure that every pixel on each diffraction pattern corresponds to the same scattering vector.
After determining the center coordinate of each diffraction pattern (i.e, the coordinate of the unscattered beam) as a function of probe position, we aligned each pattern with subpixel accuracy to the center and resampled with bilinear interpolation.
Each of the five datasets were aligned such that the unscattered beam was centered for all probe positions.
These steps were performed using the calibration tools implemented in \texttt{py4DSTEM}~\citep{{savitzky2021py4dstem}}, with more detail in Sec. \ref{sec:centering}
Finding the center beam can be challenging especially in a thick sample, where the brightest reflection is not necessarily the central spot, as exemplified by the K2-IS data shown in Fig.~\ref{fig_si:K2}.
In this case, including a vacuum region in the scan or recording a subsequent image of the probe over vacuum with the same diffraction shift alignment can be helpful. 
If one does not adjust the unscattered beam with diffraction shift, the position of the center is a known value according to the calibration as shown in Fig.~\ref{fig_si:Rot_cal} and this position value can be used as a first guess in the centering step.

Next, we calculated virtual dark-field images for each scan,  and estimated the relative shifts in real space using cross correlation. 
For samples with fewer features, such as the SiGe standard, cross-correlation can be challenging, and it was often more effective to align tilted datasets to each other than to the centered one.
Thus only one tilted dark-field image needed to be aligned to the center tilted image, which could be done automatically with the most similar tilted image or even manually if needed.
The five datasets were then aligned in real space pixel-wise by rolling the datacube and then cropping to the region that contained data from all five scans.
Fig.~\ref{fig_si:real_space_shifts} shows five of the tilts overlaid on top of each other. 
The datasets were only aligned with pixel-wise precision to reduce interpolation errors from multiple resampling of datasets. 
However, this reduces the real space resolution of the data, which could be limiting for some applications, in which case subpixel alignment would be appropriate.

We calculated strain and orientation maps using the workflows in \texttt{py4DSTEM}. 
The ACOM analysis of the MAPED dataset was performed using the sparse correlation matching method described in~\cite{ophus2022automated}.
The excitation error was modified to include a precession angle of 1$^\circ$, following previous work ~\citep{diebold2025template}.
After determining the best-fit orientation, we calculated the strain for each probe position using the following procedure.
For each probe position, we simulated a reference diffraction pattern.
Next, each simulated Bragg disk was paired with the closest experimentally measured Bragg disk, or removed if no disks were found within the radial threshold (set to 0.04 \AA{}$^{-1}$).
We then used weighted least squares to calculate the best fit affine transformation matrix which deforms the reference pattern to match the experimental pattern.
Finally, we plotted the infinitesimal strain from the transformation matrix for each probe position in the probe scanning coordinate system.

\subsection{Simulations}

We performed diffraction simulations  in \texttt{py4DSTEM} using the dynamical diffraction module with the Weickenmeier-Kohl plus core and phonon absorption parameters at 300~kV with a maximum scattering angle of 4 \AA{}$^{-1}$. 
Instead of tilting the beam, the incident zone axis for the simulations was tilted accordingly. 
We simulate silicon with the [011] aligned with the electron beam, matching the experimental data in Fig.~\ref{fig2} for silicon.
Fig.~\ref{fig4} shows the result of our calculations.
We include one simulation at zero beam tilt, and N-1 equally spaced radially at 17 mrad. 
We performed simulations for sample thicknesses from 10 to 300 nm in 10 nm steps, using 1, 5, 9, 17, 33, 65, 129, 257, and 513 tilts.

\section{Results}

\subsection{Strain mapping}

\begin{figure*}[ht]
\centering
\includegraphics[width = 0.95\textwidth]{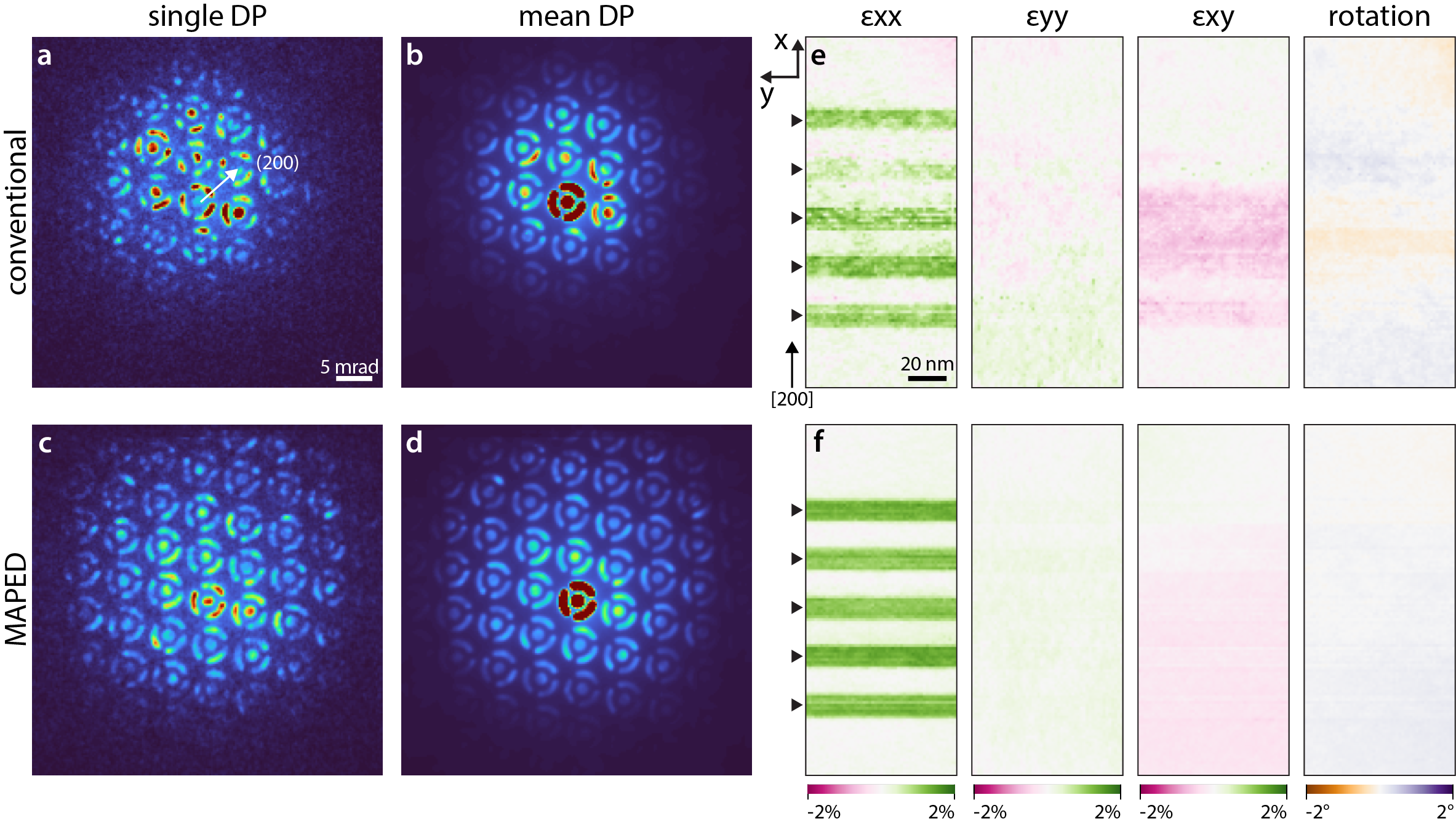}
  \caption{(a) Single and (b) mean diffraction pattern from conventional experiment and corresponding (c--d) MAPED experiment. Strain maps from the (e) conventional and (f) MAPED experiment. Black arrows indicate location of alloy regions, where we expect large $\epsilon_{xx}$. Compared to MAPED, the conventional experiments show more error in the strain maps.
}
  \label{fig2}
\end{figure*}

Fig.~\ref{fig2} shows a comparison between strain mapping with the MAPED method and a conventional 4D-STEM experiment on the SiGe sample. In this sample, we expect lattice dilatation in the alloy layers compared to the pure silicon, as germanium has a larger lattice parameter than silicon (5.67 \AA~versus 5.43 \AA)~\citep{jain2013commentary}.
Because the y-direction is constrained due to the epitaxial growth, we expect the strain to be localized to the x-direction and the expansion of the lattice should create a tensile (positive) strain along that direction.

These datasets were collected with a bullseye aperture to improve Bragg disk detection as described by ~\cite{zeltmann2020patterned}.
Fig.~\ref{fig2}a--b shows a single and mean diffraction pattern from a conventional experiment and Fig.~\ref{fig2}c--d plots the corresponding MAPED patterns. 
The intensities of higher order Bragg reflections are clearly reduced in Fig.~\ref{fig2}a--b, as a consequence of the curvature of the Ewald sphere. 
For the same dose, the data acquired with MAPED (Fig.~\ref{fig2}c--d) shows that higher order Bragg reflections, which are excited upon tilting the beam. 

Fig.~\ref{fig2}e--f show the strain maps from the same field of view with and without MAPED, and the corresponding dark-field image is shown in Fig.~\ref{fig_si:strain_df}. 
A similar SiGe alloy sample has often been used as a benchmark for measuring strain with S/TEM techniques~\citep{rouviere2013improved, mahr2021accurate, munshi2022disentangling}.
Here $\epsilon_{xx}$ corresponds with the x direction that is vertical to the page as indicated by the arrows to the left of the strain maps.
Based on previous studies in the literature, we expect the strain to be approximately 1.2\%~\citep{rouviere2013improved, munshi2022disentangling}.
The dataset collected with MAPED is much closer to the predicted strain values, as shown in Fig.\ref{fig2} and the line profiles in Fig.~\ref{fig_si:line_profile}. The $\epsilon_{xx}$ strain is close to uniform between and across different sets of alloy layers and shows tension due to the expanded lattice parameter. 
There are some additional features in the MAPED $\epsilon_{yy}$, $\epsilon_{xy}$, and $\theta$ strain maps, including shear strain across the field of view and small differences in strain between SiGe layers; some of these could be real features of our sample.
 
Nonetheless, in comparison, the strain maps for the conventional scans show significant errors, including non-uniform strain between the sets of alloy layers, which is not expected. 
These errors are underscored by the histograms of strain from a control (non-alloy region) as shown in Fig.~\ref{fig_si:strain_hist}. 
The distribution of strain intensities from an ideally control unstrained region is much higher without the MAPED method.

There are multiple reasons why the strain mapping in the conventional experiment would be less reliable. 
First, a single tilt experiment is more susceptible to intensity variations in the disks leading to less accurate disk detection; comparing Fig.~\ref{fig2}a\&c, it is clear how MAPED leads to more uniform intensity across disks. 
Second, in the MAPED experiment, disks at  higher scattering angles are more strongly excited and thus more clearly identified. 
The strain algorithm fits higher-order peaks in addition to first order reflections for more robust strain measurements, so the extra signal at high scattering angles can improve calculations. 
The strain measurements in both datasets could be further enhanced through advanced processing such as described by ~\cite{munshi2022disentangling}. 
Ultimately, the combination of sequential tilting and bullseye apertures leads to highly accurate strain measurements.

\subsection{Orientation mapping}

\begin{figure*}[ht]
\centering
\includegraphics[width = 0.95\textwidth]{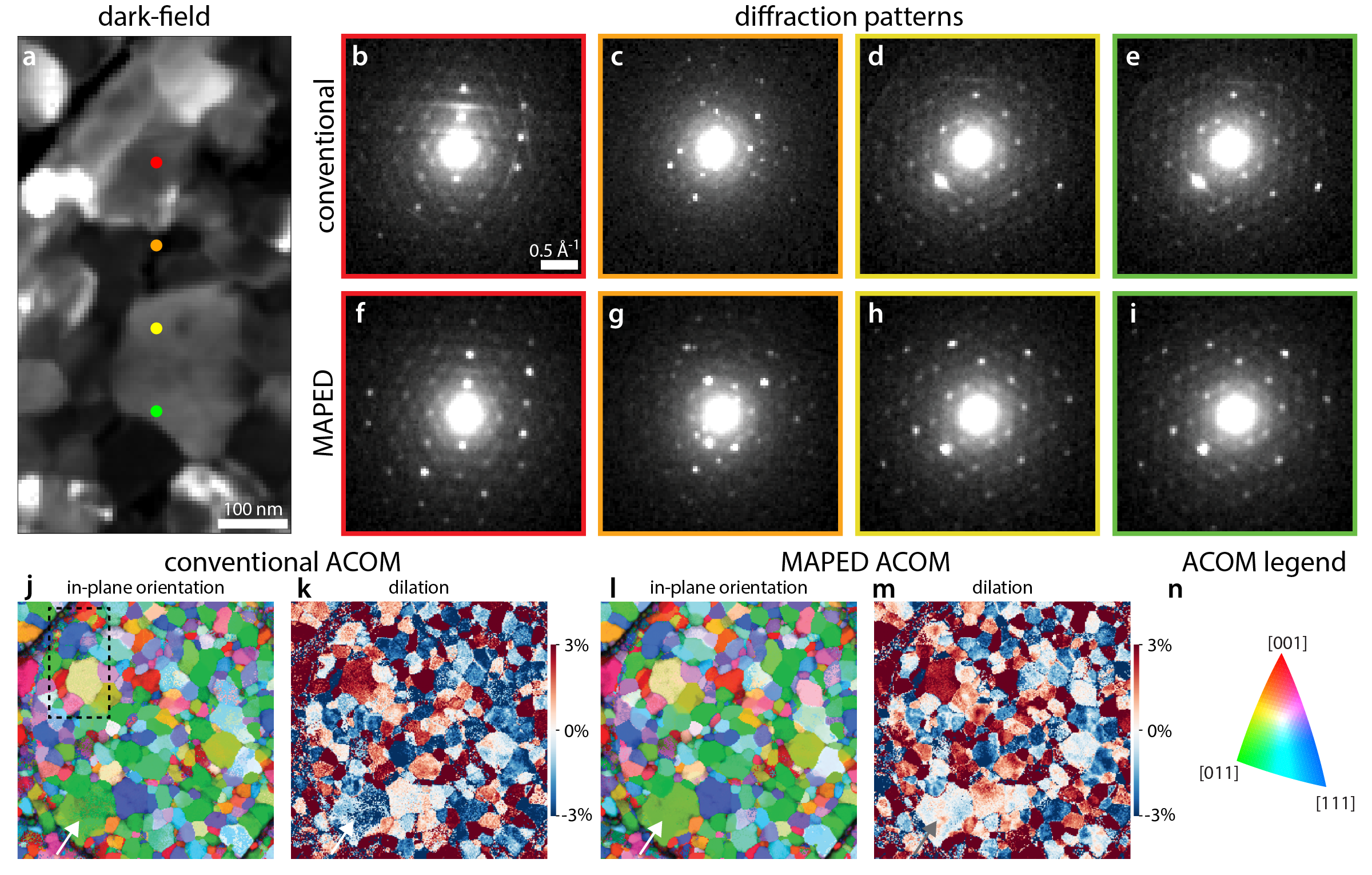}
  \caption{(a) dark-field image of polycrystalline aluminum film. (b--e) Conventional diffraction patterns from 4 spots in (a). (f--i) Corresponding MAPED diffraction patterns show enhanced peaks at high scattering angles and reduction of Kikuchi diffraction lines.
  ACOM maps and dilation ($\epsilon_{xx}$ + $\epsilon_{yy}$) from (j--k) conventional and (l--m) MAPED data with (n) legend}. Dark-field region in (a) marked with rectangle in (j). 
  \label{fig3}
\end{figure*}

To further explore the application of our MAPED  method to study material structure, we performed ACOM analysis on a polycrystalline aluminum film (Fig.~\ref{fig3}).
From the dark-field image (Fig.~\ref{fig3}a) we plot four diffraction patterns from the same position. 
The top row presents conventional STEM diffraction patterns, while the bottom row shows data from the same positions but collected with our MAPED method. 
As with the data in Fig.~\ref{fig2}, incorporating more tilts leads to diffraction information at higher scattering angles. 
Moreover, the MAPED method suppressed dynamical artifacts such as Kikuchi diffraction, which is seen comparing Fig.~\ref{fig3}b\&f.

The orientation maps and strain calculations from the ACOM maps are shown in Figs.~\ref{fig3}j--m. 
The conventional diffraction data approach yields high-quality orientation matches. 
The in-plane rotation maps plus the out-of-plane orientation (Fig.~\ref{fig_si:acom_conventional}) show most of the grains within the film are approximately aligned with the [111] parallel to the beam with in-plane rotation between grains. 
Finally, using the structure file as a reference, the strain in each grain is calculated, with the dilation ($\epsilon_{xx}$ + $\epsilon_{yy}$) plotted in Fig.~\ref{fig3}k.
The full strain maps are shown in Fig.~\ref{fig_si:acom_strain_conventional}.

The same experiment performed on the identical field of view using MAPED yields qualitatively similar results, but with significantly improved signal-to-noise in both the ACOM maps and strain calculations (Figs.~\ref{fig3}l--m, \ref{fig_si:acom_precession}, and \ref{fig_si:acom_strain_precession}).  
The improvement is especially evident in the strain maps, where the MAPED data produces far more physical results. 
This trend is evident, for example, in the large grain at the bottom left of the field, highlighted by the arrow in Figs.~\ref{fig3}j--m. While the polycrystalline aluminum film studied here serves as a model system to compare ACOM with and without MAPED, we anticipate that this improvement in ACOM matching will be even more significant in challenging samples.

\subsection{Quantification of intensities}\label{sec:quant}

\begin{figure*}[ht]
\centering
\includegraphics[width = 0.95\textwidth]{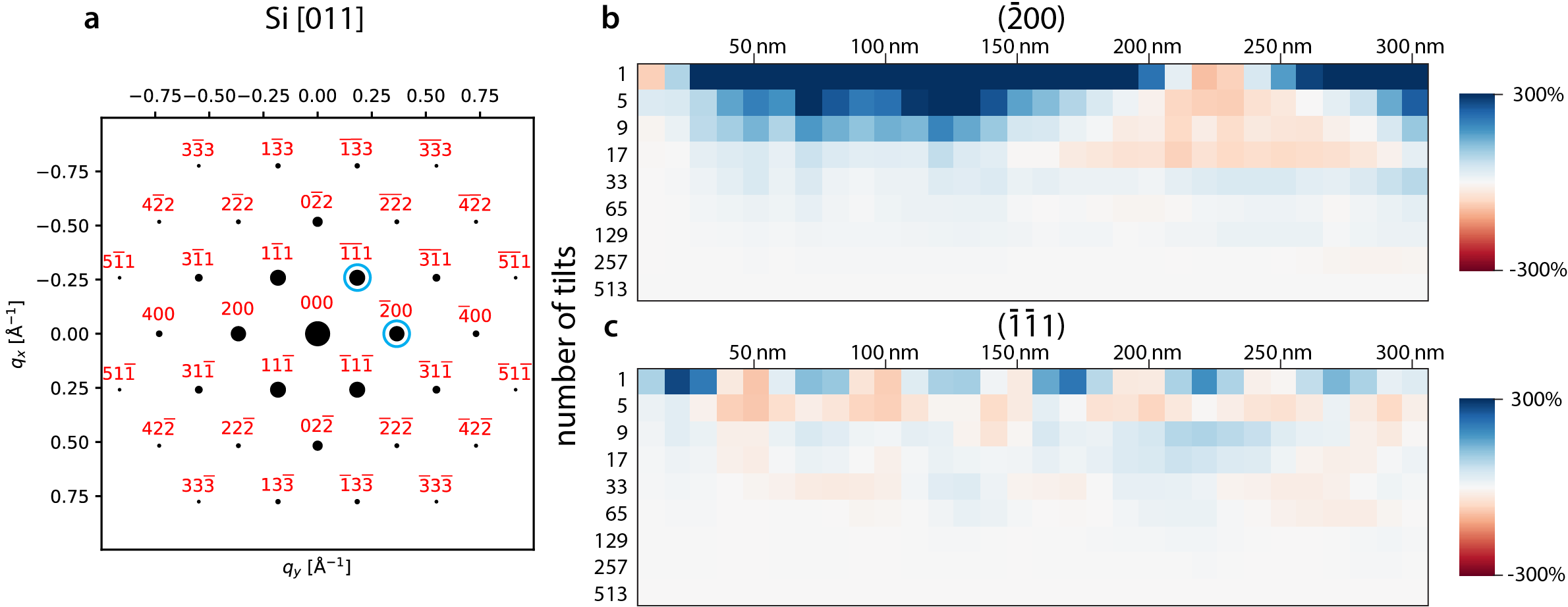}
  \caption{Dynamical diffraction simulations to explore impact of few tilt precession on intensity quantification. (a) Si [011] diffraction pattern. The ($\overline{2}00$) and ($\overline{11}1$) intensity values are shown in (b) and (c) as a function of thickness and number of tilts. Intensity values normalized to show percent difference from the intensity value with the highest number of tilts.}
  \label{fig4}
\end{figure*}

Although our MAPED method can increase the signal from high scattering angles as shown in Figs.~\ref{fig2}\&\ref{fig3}, there is an anisotropic nature to the method, which non-uniformly boosts the signal in the diffraction patterns. 
For example, consider the five diffraction patterns from the silicon sample shown in Fig.~\ref{fig1}a. 
The tilt of the Ewald sphere enhances the signal of some disks more than others.

To better quantify the impact our MAPED method has on diffraction pattern intensities, we performed dynamical Bloch wave simulations in \texttt{py4DSTEM}.
In Fig.~\ref{fig4}, we plot the the percent difference in intensity in each reflection as compared to the maximum number of tilts (bottom row).
As more tilts are included, the differences decrease, and the magnitude of this result will depend on sample thickness, tilt angle, and the number of tilts used.
One can imagine how continuously precessing systems, which more finely sample the rotation space, facilitate predictable intensity values and more complete sampling of diffraction space. 

It is also important to note how the intensity of these values varies with thickness, which is most easily observed in the intensity plots shown in Fig.~\ref{fig_si:sim}.
When only one tilt is acquired, as in conventional experiments, the intensity of the ($\overline{11}1$) reflection oscillates with thickness, a well-known feature of dynamical diffraction; this is the first row in the plot in Fig.~\ref{fig_si:sim}.
With just four additional tilts, as in the MAPED data, these thickness oscillations are significantly decreased and less sensitive to thickness, as with conventional precession experiments~\citep{midgley2015precession}. 
The ($\overline{2}00$) reflection is a kinematically forbidden reflection, and it is expected that with precession that this intensity will be significantly decreased and less sensitive to thickness~\citep{vincent1994double, ciston2008quantitative,eggeman2010precession}.  Our simulated MAPED data is in agreement, as including even four additional tilts suppresses the intensity of these reflections. 
Overall these results highlight how MAPED experiments make intensities more kinematic-like in thicker samples.

\section{Discussion}
Precession is well-understood to improve diffraction data by reducing dynamical diffraction artifacts and collecting higher scattering angles. 
Here, we have highlighted how collecting data at a few tilts can lead to high-quality diffraction data with many of the benefits of integrated precession experiments. 
The flexibility in choice of cameras and microscopes to collect precession data also poses a significant advantage for 4D-STEM experiments. 

Our MAPED method helps with one of the challenges in aligning precession tools for non-integrated cameras.
Precession systems that are integrated with a camera address a core challenge of precession 4D-STEM: a precession-induced ``descan," which is a shift of unscattered beam in the back focal plane as a consequence of tilting the incident illumination.   
This precession-descan is in addition to the  common descan observed when an electron beam is rastered across a specimen, and it is elegantly handled by automated routines in the integrated systems. 
If one does not have a commercial precession integrated with the camera of choice, it is possible to use a camera that is not integrated by having one scan generator do the X-Y scanning in synchronization with the camera and a second scan generator do precession and precession-descan correction.  
This mode does not have automated routines, so fine tuning of the precession-descan correction with the second scan generator is challenging. 
Without perfect alignment, the Bragg disks can become blurred due to precession-descan, leaving no ability to correct computationally after acquisition. 
However, in our MAPED method, we are able to preserve sharp features in our diffraction disks, even in the challenging case of bullseye apertures as shown in Fig.~\ref{fig2}, because both types of descan are addressed in post-processing.

\begin{figure*}[ht]
\centering
\includegraphics[width = 0.95\textwidth]{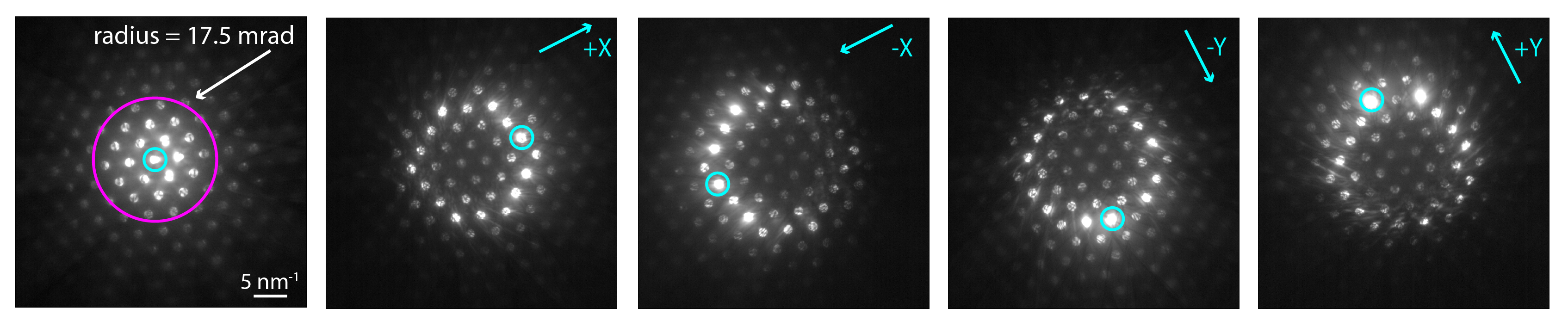}
  \caption{Example of beam tilting in an uncorrected microscope. Here, the rotation center is use to tilt the beam by ~1$^\circ$ (17.5 mrad) shown by the magenta circle on a silicon [110] sample. Diffraction from the sample is used as a calibration for reciprocal sampling, which is used to determine the rotation center value for a desired beam tilt. The unscattered beam is circled in blue.}
  \label{fig5}
\end{figure*}

The previous sections show data collected on an aberration corrected microscope in nanoprobe mode, where the beam tilt is controlled by the probe corrector software.  
In an uncorrected TEM, the rotation center can be used to tilt the beam. 
The practical result is the same -- the probe is tilted with respect to the optic axis.
In Fig.~\ref{fig5} we show the four tilts collected on an uncorrected Thermo Fisher Scientific ThemIS using the rotation center alignment feature to tilt the beam. 
In this experiment, we used the lattice spacing from the diffraction to approximate the tilt of the beam with rotation center.

To implement MAPED across these microscopes, our experiments have also relied on calibrations.
Because the tilting of the beam is done manually (either through rotation center or the corrector software), it is important to calculate the approximate beam tilt.
We showed two methods for calibrating tilts, either with the semi-convergence angle (Fig.~\ref{fig_si:team_i_cali}) or the lattice itself (Fig.~\ref{fig5}).
The tilts need to be calibrated for each microscope and accelerating voltage, and we have found these values to be stable for subsequent sessions on different days. 
Note these are approximate tilts; there are additional factors that could be considered for more accurate calibration including aberrations and beam shift~\citep{barnard2017high, nordahl2023correcting, flathmann2025sequential}. 
Nonetheless, an approximate calibration is sufficient for determining proper data collection parameters for strain and orientation mapping. 

An additional consideration is the practicality of acquiring these few tilt 4D-STEM scans.
Collecting and processing MAPED datasets requires additional microscope time and computational resources over conventional experiments. 
However, automated precession pipelines through vendor hardware also require additional overhead and ultimately require similar collection times when competing with fast cameras. 
An additional practical advantage of the described sequential approach is that the microscope stays in STEM mode, as compared to commercial systems, which usually use a converged probe in TEM mode. 
Collected data in STEM mode allows easier correlative acquisition of other microscopy signals, including annular dark-field STEM and STEM spectroscopy for multimodal analysis. 

Despite these benefits there are limitations for the MAPED precession approach. 
As studied in Sec.~\ref{sec:quant}, especially for thicker samples, the intensities depend on tilts.
The extent to which this impacts measurement of sample properties depends strongly on the type of reconstruction.
The uniformity in intensities has very little impact on strain measurements. 
For many ACOM experiments, the intensity of reflections is not weighted strongly when making matches from an existing library of structures. 
In fact this is the default in \texttt{py4DSTEM}.
However, when mapping multiple phases from the same field of view, intensity values can be helpful for separating similar phases and for probing subtle structural changes due to defects~\citep{diebold2025template, correa2024high}. 
It would be possible to incorporate the asymmetry in intensities that comes from discrete instead of continuous tilts in ACOM simulations for better matching.
The practical benefits and limitations of the MAPED approach, as compared to conventional precession, could be more directly compared on a microscope with built-in precession.

To improve reliability and robustness of MAPED and extend it to a larger tilt series for more accurate quantification, we anticipate the need for automation of the data collection.
This requires scripting to automatically tilt the beam, re-center the unscattered beam if necessary, 
and acquire and save the datasets with sufficient metadata. 
Acquiring sequential precession datasets increases the memory required to store data, underscoring the importance of compression schemes for storing data. 
Lastly, the MAPED pipeline would be improved through more automated methods for data processing, which will become increasingly important if the number of tilted scans increases.

\section{Conclusion}

We have presented MAPED: a sequential, few-tilt approach to collecting precession electron diffraction data. 
This method enables improved measurement of material structure and properties compared to single-tilt 4D-STEM experiments, as demonstrated by examples of strain mapping in a SiGe alloy and automated phase identification in a polycrystalline aluminum film. 
We also used dynamical diffraction simulations to assess the limitations of this approach for accurate intensity quantification, particularly in thicker samples.

While collecting multiple scans from the same field of view is more time-consuming and data-intensive than conventional 4D-STEM, alignment and descan correction on integrated precession systems can be similarly complex, especially when non-integrated cameras are used. 
Our results show that this sequential tilt approach is compatible with a variety of microscopes and detectors, offering more experimental flexibility than conventional hardware-integrated precession systems.

Analysis and processing of MAPED data are feasible using open-source tools.
We expect this method to expand access to precession-enhanced experiments, particularly for challenging samples where conventional techniques may fail. 
Although we used only five tilts here, our simulations indicate that using more tilts can improve results. 
In future work, automation of beam tilting and data collection will be essential to scale this sequential precession approach to larger tilt series and make it more broadly accessible.

\section{Competing interests}
No competing interest is declared.

\section{Data availability}
The code is available on \href{https://github.com/smribet/few_tilt_precession}{github}.

\section{Author contributions statement}

R.D., C.O., and S.M.R. conceived of the experiments,  S.M.R. and K.C.B. conducted the experiments, all authors analyzed the results,  S.M.R. and K.C.B. wrote the manuscript draft, all authors reviewed the manuscript.

\section{Acknowledgments}
Work at the Molecular Foundry was supported by the Office of Science, Office of Basic Energy Sciences, of the U.S. Department of Energy under Contract No. DE-AC02-05CH11231. S.M.R., K.C.B. and C.O. acknowledge support from the U.S. Department of Energy Early Career Research Award program. We thank Bin Jiang of Thermo Fisher Scientific for helpful consultation.

\newpage

\bibliographystyle{plainnat}
\bibliography{reference}

\newpage
\clearpage

\setcounter{section}{0}
\renewcommand\thesection{S\arabic{section}.} 

\setcounter{page}{1}
\renewcommand\thepage{S.\arabic{page}}      

\setcounter{figure}{0}  
\renewcommand\thefigure{S.\arabic{figure}}  

\setcounter{equation}{0}
\renewcommand{\theequation}{S\arabic{equation}}

\onecolumn
\begin{center}
\textbf{{\Large Multi-angle precession electron diffraction (MAPED): a versatile approach to 4D-STEM precession}}

\end{center}
{\large Stephanie M. Ribet$^{1,\dagger,*}$, Rohan Dhall$^{1,\dagger}$, Colin Ophus$^{1,*}$, Karen C. Bustillo$^{1,*}$}

~\hspace{-2em} $^{1}$ National Center for Electron Microscopy, Molecular Foundry, Lawrence Berkeley National Laboratory, Berkeley, 94720, CA, USA

~\hspace{-2em} $^{2}$ Department of Materials Science and Engineering, Stanford University, Palo Alto, 94305, CA, USA

~\hspace{-2em} $^{\dagger}$These authors contributed equally to this work

~\hspace{-2em} $^{*}$Authors to whom correspondence should be addressed: sribet@lbl.gov, kbustillo@lbl.gov

\section{S1. Reciprocal space centering}\label{sec:centering}
Centering and aligning a 4D-STEM dataset in reciprocal space can be challenging, especially with thicker samples or unconventional apertures.
To address this challenge, there are several centering routines implemented in  \texttt{py4DSTEM}, which were all leveraged in this work. 
The simplest centering method assumes the brightest pixel is the center spot in each diffraction pattern, and this approach was leveraged for the centering in Fig.~\ref{fig3}.
This method is only valid in thin samples.
To find the center, each diffraction pattern is gaussian filtered and the location of the brightest pixel is identified. 
The center is further refined with a center of mass routine. 
For thicker samples or data acquired with a bullseye aperture (such as that shown in Fig.~\ref{fig2} or Fig.~\ref{fig5}), a more complicated method is needed to find the center.
Disk detection is used to identify the spots in each diffraction pattern, which is important for precisely locating the position of each Bragg disk. 
The 4D-STEM data is converted to BraggVectors, and then a variety of centering routines can be applied.
A center guess can be supplied by the user, and the center can be found by the following scoring metrics (1) distance from center guess, (2) intensity, (3) intensity weighted distance from the central peak. 
For Figs.~\ref{fig2} \& ~\ref{fig5}, the distance from the center guess was used. 
After the center is measured at each probe position with either method, it is fit to a function to include only low frequency shifts due to descan and not high frequency shifts due to sample potential. In this case a plane was used.

\newpage
\section{S2. Supplemental Figures}

\begin{figure*}[ht]
\centering
\includegraphics[width = 0.95\textwidth]{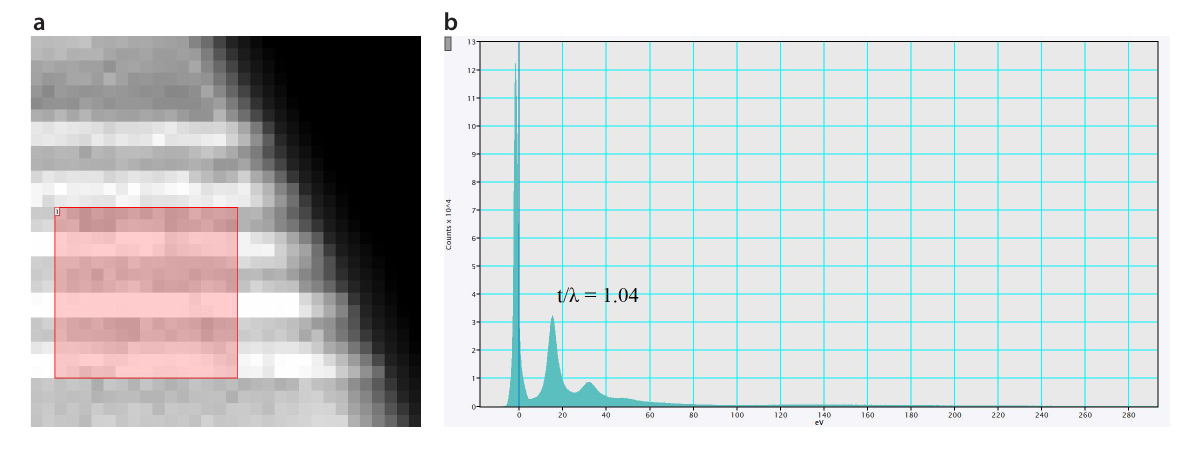}
  \caption{Thickness measurement using EELS on K3 camera with MAG*I*CAL sample (a) Dark-field image with red area the region of interest for thickness measurement and (b) EELS spectrum in low-loss, t/$\lambda$ calculated to be about 1 for the sample in Fig.~\ref{fig2}
  }
  \label{fig_si:eels}
\end{figure*}

\begin{figure*}[ht]
\centering
\includegraphics[width = 0.95\textwidth]{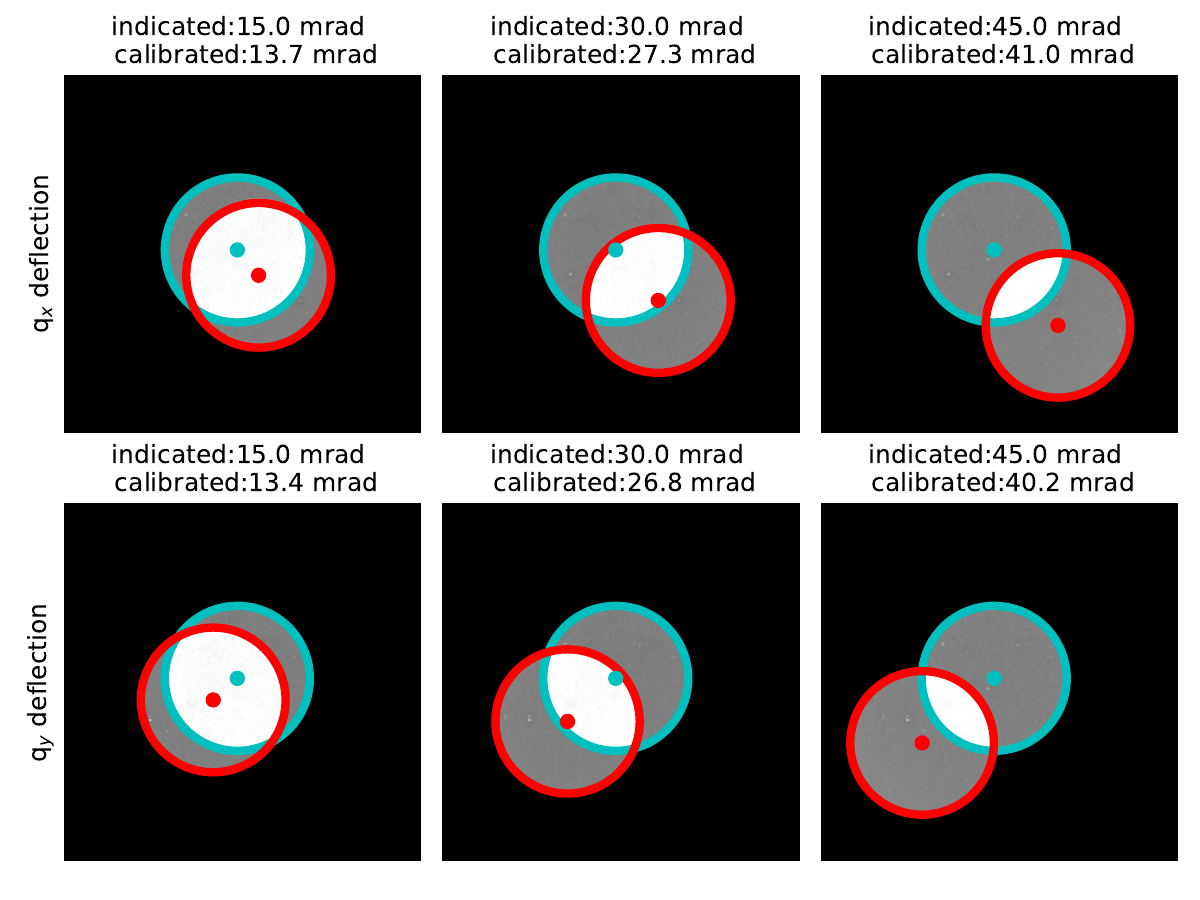}
  \caption{Calibration of deflection of beam using a large convergence probe (30 mrad). The beam was tilted with the corrector software. The calibrated value is about 10\% smaller than the expected value.  This calibration does not include the beam shift.
  }
  \label{fig_si:team_i_cali}
\end{figure*}

\begin{figure*}[ht]
\centering
\includegraphics[width = 0.95\textwidth]{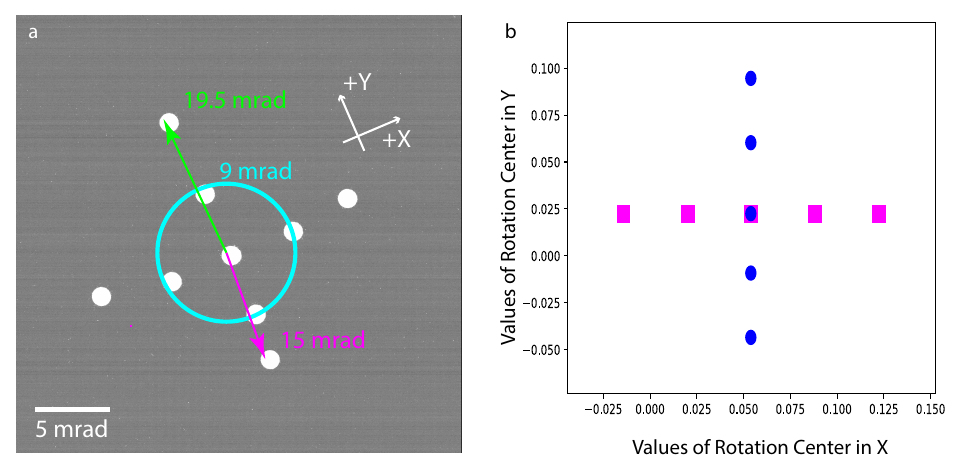}
  \caption{Rotation Center Calibration on Thermo Fisher ThemIS TEM a)  Sum of nine images of the unscattered beam in vacuum.  The initial values were set based on a 9 mrad circle in turquoise.  To investigate the linearity of the alignment we added the same amount to each value to target an 18 mrad circle; the arrows in magenta and green show that the actual tilt for a given change in quantity ranged from 15 to 19.5 mrad. Knowledge of the approximate locations for the tilt allows one to provide starting guesses for initial centering of the diffraction patterns. b) The values for the Rotation Center X and Y corresponding to the 9 tilts in a).
  }
  \label{fig_si:Rot_cal}
\end{figure*}

\begin{figure*}[ht]
\centering
\includegraphics[width = 0.95\textwidth]{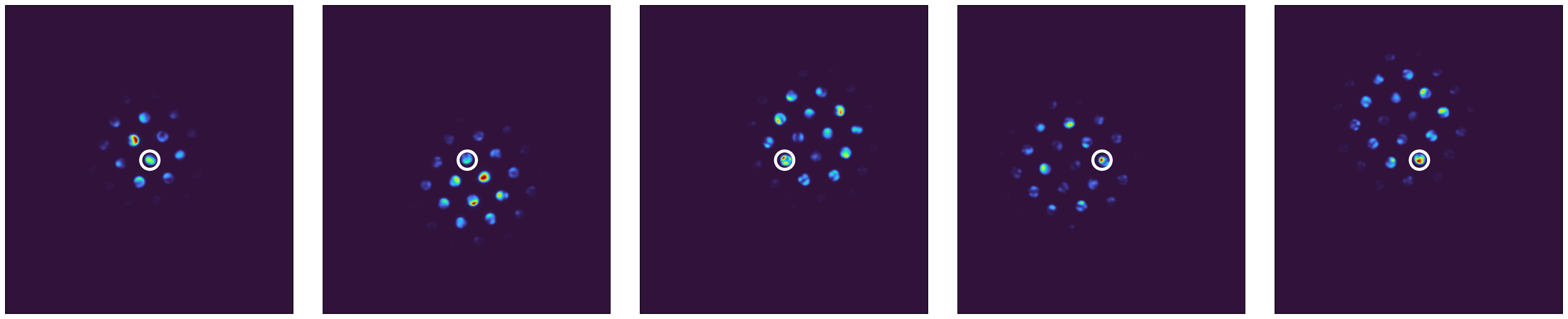}
  \caption{Five tilt series from a thick region of of a Si sample acquired on the K2-IS camera. Due to dynamical diffraction in a thick sample, the brightest spot is not the central beam. The unscattered beam is circled. Here, prior calibrations were used to manually correct central spot for scan alignment.
  }
  \label{fig_si:K2}
\end{figure*}

\begin{figure*}[ht]
\centering
\includegraphics[width = 0.75\textwidth]{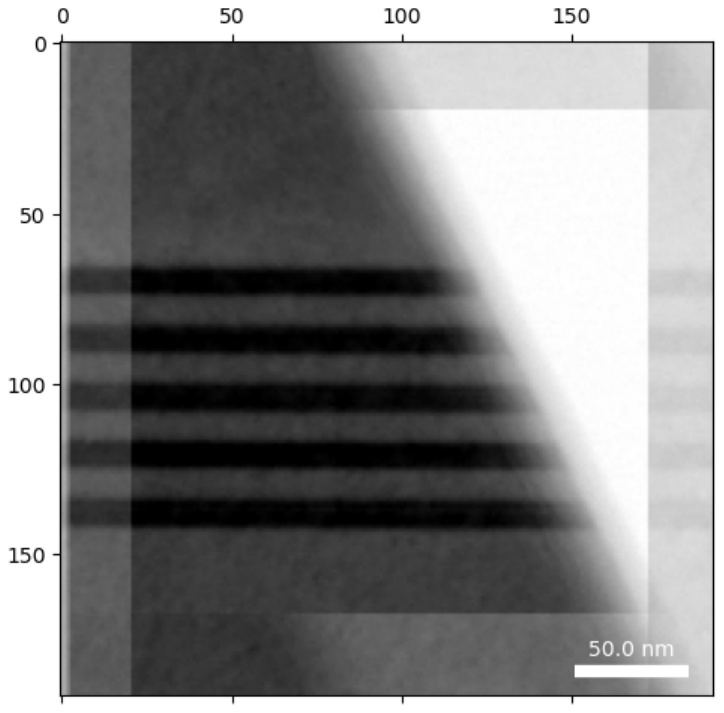}
  \caption{Real space alignment of sample from Fig.~\ref{fig2}. The image shift is due to the beam shift with tilting and is not included in the calibration of the tilt.  Neither the region-of-interest box or the sample were moved between different tilts.
  }
  \label{fig_si:real_space_shifts}
\end{figure*}

\begin{figure*}[ht]
\centering
\includegraphics[width = 0.25\textwidth]{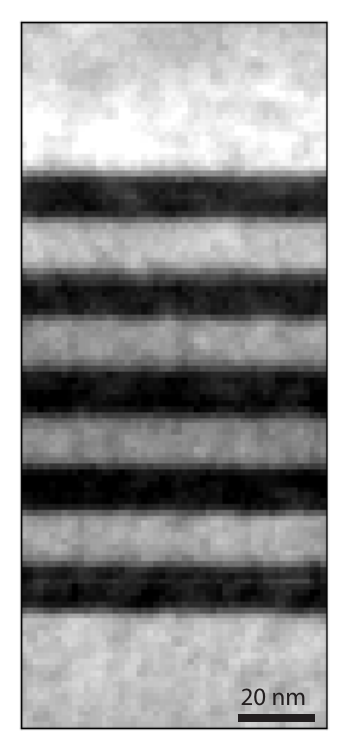}
  \caption{Dark-field image of field-of-view in Fig.~\ref{fig2}
  }
  \label{fig_si:strain_df}
\end{figure*}

\begin{figure}
    \centering
    \includegraphics[width=0.5\linewidth]{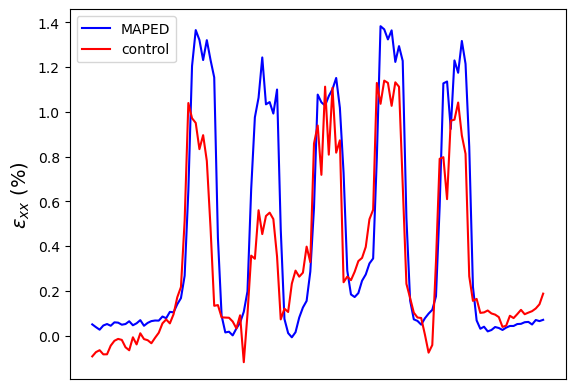}
    \caption{$\epsilon_{xx}$ strain profiles from Fig.~\ref{fig2} averaged along the y-direction.}
    \label{fig_si:line_profile}
\end{figure}

\begin{figure*}[ht]
\centering
\includegraphics[width = 0.95\textwidth]{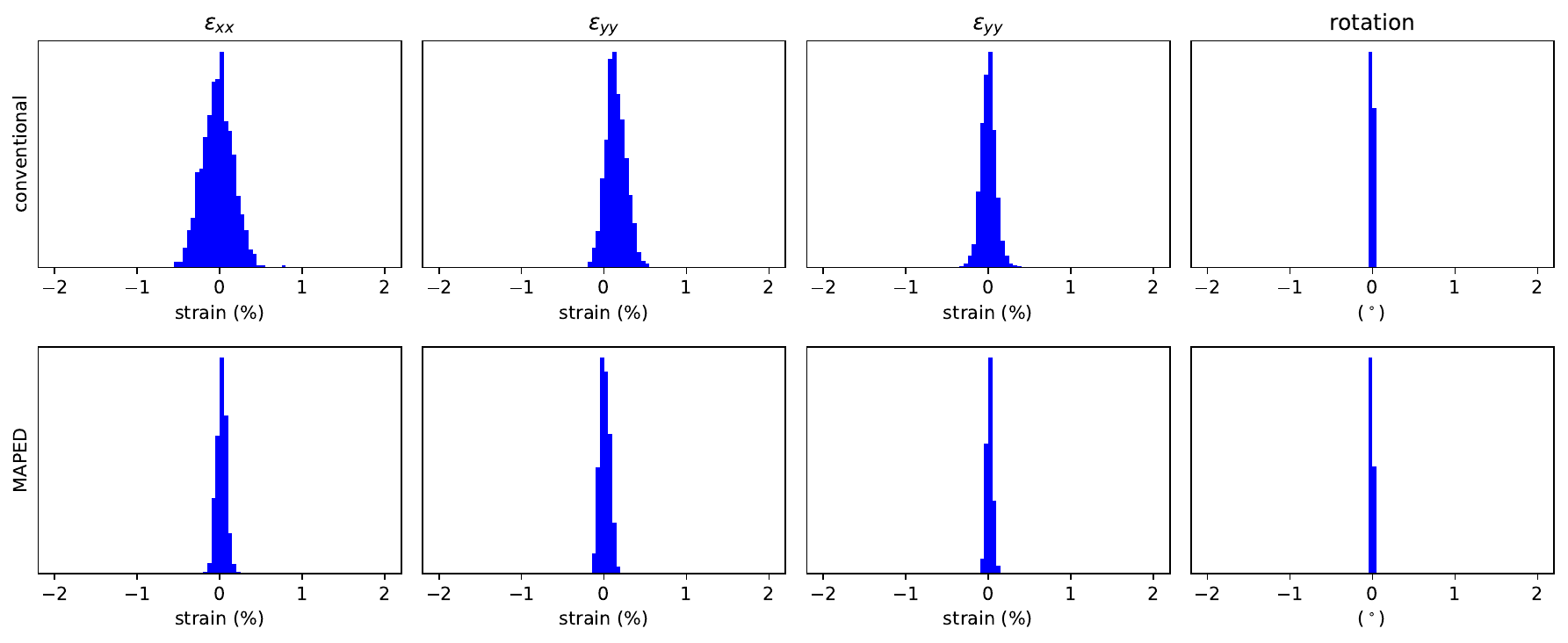}
  \caption{A histogram of strain values from a control region next to the field of view from Fig.~\ref{fig2} acquired with the same parameters. This region is away from the alloy and should be strain free. The MAPED method produces fewer error as shown by the tighter distribution in the histogram. 
  }
  \label{fig_si:strain_hist}
\end{figure*}

\begin{figure*}[ht]
\centering
\includegraphics[width = 0.95\textwidth]{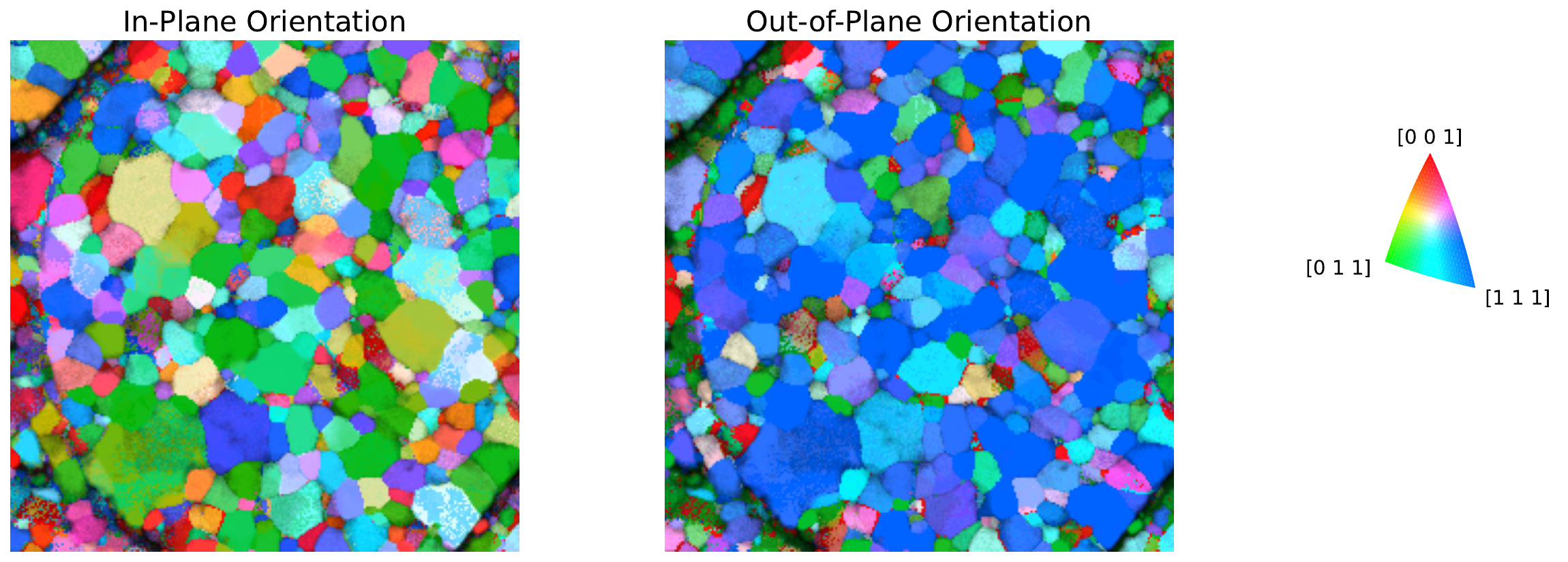}
  \caption{Full ACOM results for conventional scan from Fig.~\ref{fig3}
  }
  \label{fig_si:acom_conventional}
\end{figure*}

\begin{figure*}[ht]
\centering
\includegraphics[width = 0.95\textwidth]{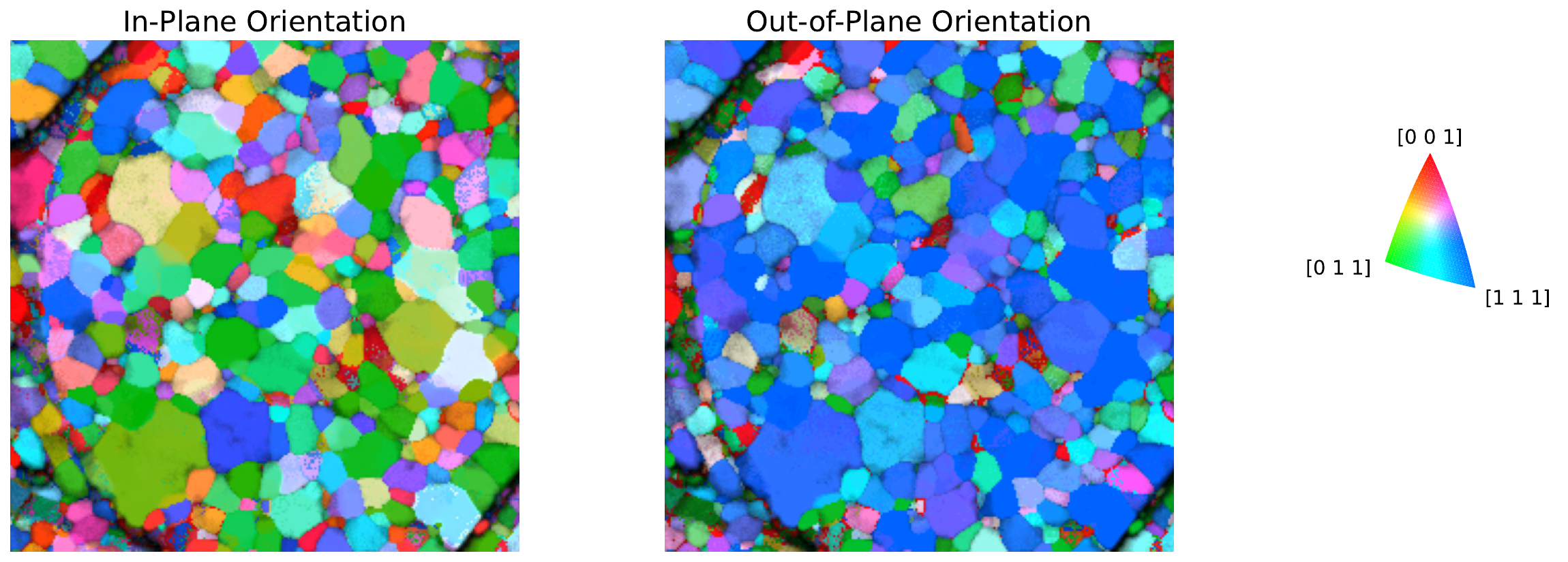}
  \caption{Full ACOM results for MAPED scan from Fig.~\ref{fig3}
  }
  \label{fig_si:acom_precession}
\end{figure*}

\begin{figure*}[ht]
\centering
\includegraphics[width = 0.95\textwidth]{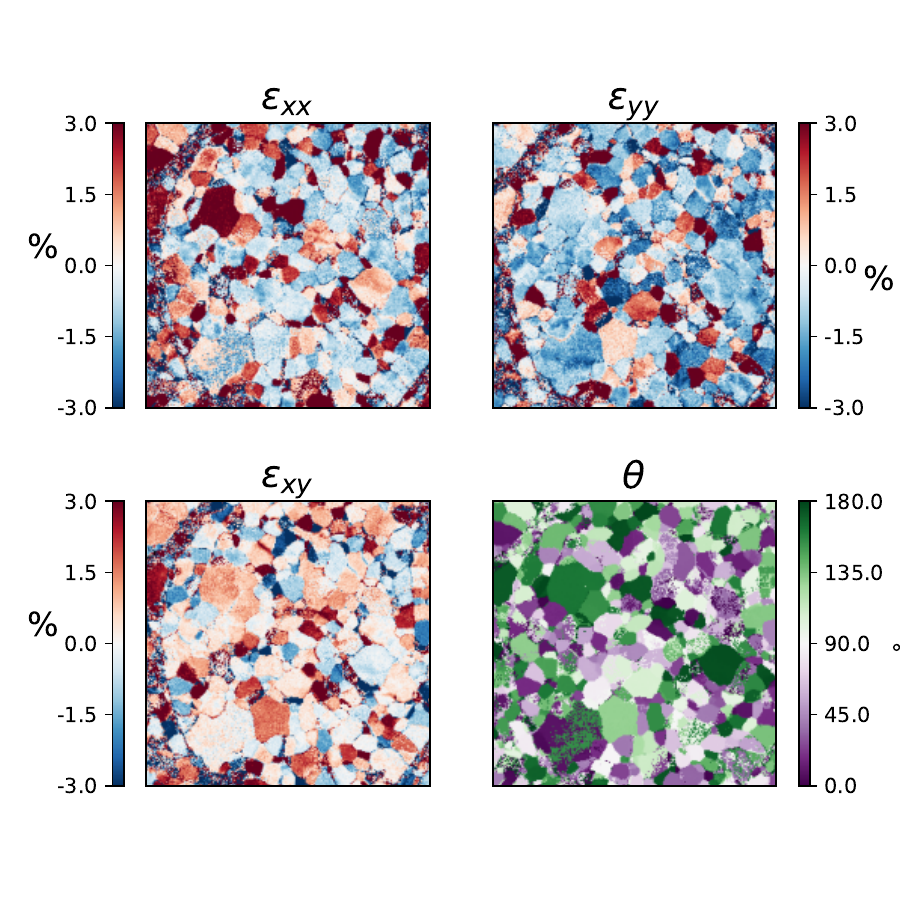}
  \caption{Full ACOM strain results for conventional scan from Fig.~\ref{fig3}
  }
  \label{fig_si:acom_strain_conventional}
\end{figure*}

\begin{figure*}[ht]
\centering
\includegraphics[width = 0.95\textwidth]{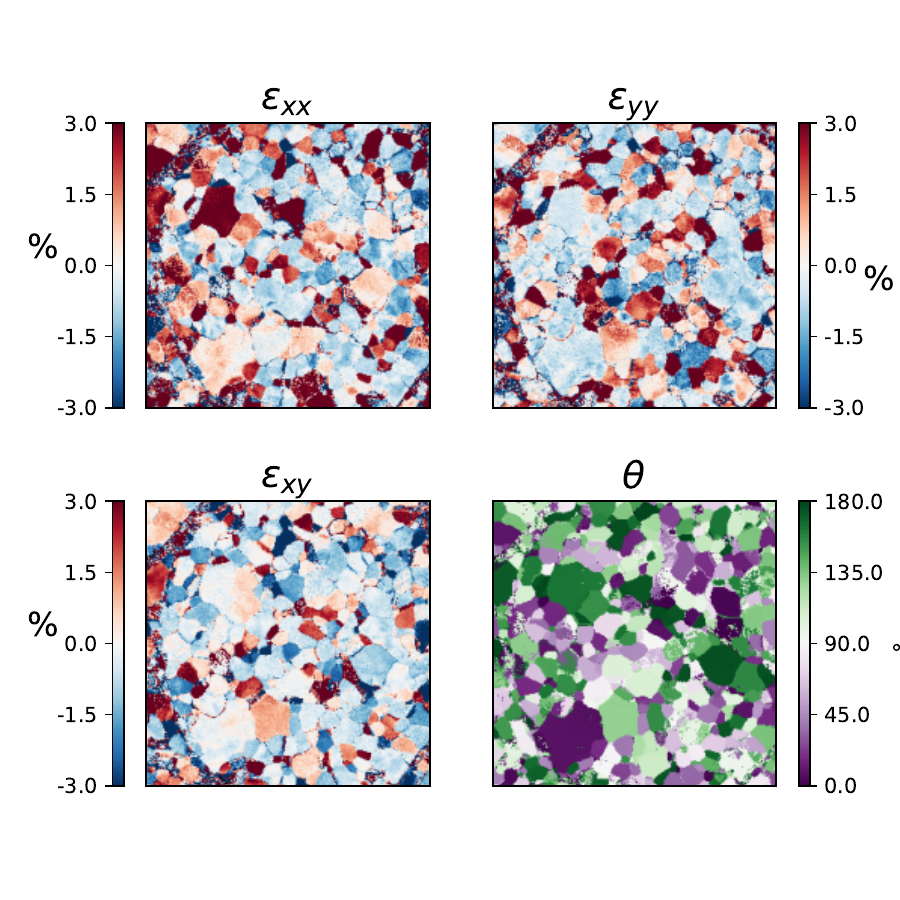}
  \caption{Full ACOM strain results for MAPED scan from Fig.~\ref{fig3}
  }\label{fig_si:acom_strain_precession}
\end{figure*}

\begin{figure*}[ht]
\centering
\includegraphics[width = 0.95\textwidth]{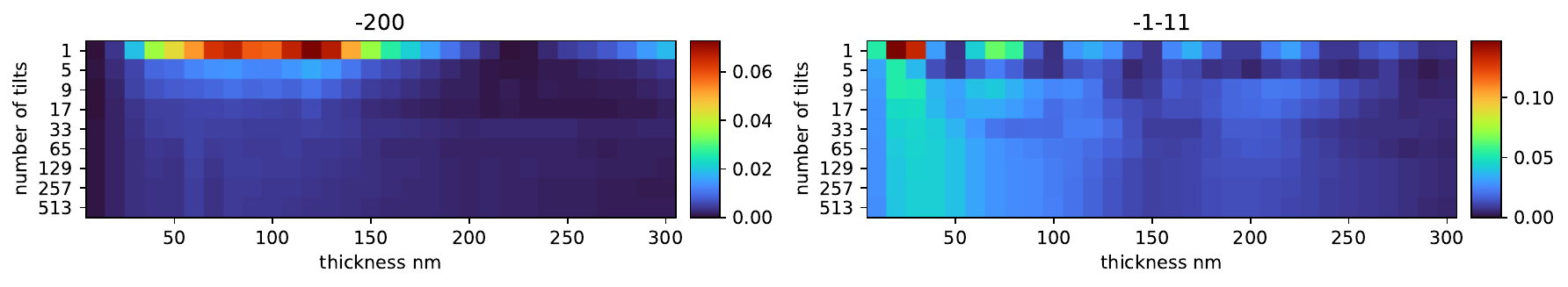}
  \caption{Comparing sampling of tilts to intensity quantification. 
  }
  \label{fig_si:sim}
\end{figure*}

\end{document}